\documentclass[manuscript]{aastex}

\usepackage{graphics,epsfig}

\slugcomment{Draft version (Marzo 2, 2010)}

\begin{document}

\shorttitle{The discovery of a molecular cavity in the direction of
Westerlund~1 region}

\shortauthors{Luna, A., Mayya, Y.~D., Carrasco, L., $\&$ Bronfman,
L.}

\title{The discovery of a molecular cavity in the Norma near arm associated to
H.E.S.S $\gamma$-ray source located in the direction of
Westerlund~1.}

\author{Abraham Luna, Y.~D. Mayya, Luis Carrasco}

\affil {Instituto Nacional de Astrof\'isica, \'Optica y Electr\'onica, Tonantzintla, Puebla, M\'exico.}

\email{aluna@inaoep.mx}

\and

\author{Leonardo Bronfman}
\affil{ Departamento de Astronom\'ia, Universidad de Chile, Casilla 36D Santiago Chile.}

\begin{abstract}
We report on the discovery of a molecular cavity in the Norma near
arm in the general direction of Westerlund 1 (Wd1), but not
associated with it. The cavity has a mean radial velocity of
$-91.5$~km\,s$^{-1}$, which differs by as much as
$\sim$40~km\,s$^{-1}$ from the mean radial velocity of the Wd1
stars. The cavity is surrounded by a fragmented molecular shell of
an outer diameter of about 100\,pc and 10$^{6}$M$_\odot$, which is
expanding at velocities of 6 to $8$~km\,s$^{-1}$. The amount of
kinetic energy involved in the expanding shell is $\sim10^{51}$ erg.
Inside this cavity the atomic HI gas surface density is also the
lowest. Structure of the extended Very High Energetic (VHE)
$\gamma$-ray emission, recently reported by the H.E.S.S.
collaboration \citep{ohm09}, coincides with the cavity. The observed
morphology suggests that the inner wall of the molecular shell is
the zone of the $\gamma$-ray emission, and not the dense gas
surrounding massive stars of Wd1 as had been speculated by the
H.E.S.S. collaboration. A likely candidate responsible for creating
the observed cavity and the $\gamma$-ray emission is the pulsar PSR
J1648-4611.

\end{abstract}


\section{Introduction}

Recently, the H.E.S.S. collaboration carried out a search for Very
High Energy (VHE, E$>$100~Gev) $\gamma$-ray emission from Galactic
young stellar clusters and found positive detections in the
direction of Westerlund 1 (Wd1) and Westerlund 2 (Wd2) (Ohm et al.
2009; Ohm09 henceforth). The motivation behind these searches comes
from the fact that the stellar clusters are potential acceleration
sites of VHE particles, since they host a variety of energetic
sources such as Supernova Remnants (SNRs) and pulsar wind nebulae.
Extended and point like emission was detected surrounding Wd2
\citep{aharonian07}, whereas only extended emission off-centered
from the cluster was detected in the case of Wd1 (Ohm09). The
emission surrounding Wd2 was associated with the cluster by
\cite{dame07}, who found structural coincidences between the CO map
and the $\gamma$-ray source. Ohm09, searched for structural
coincidences between the $\gamma$-ray and HI maps of Wd1, and
suggested a possible association of Wd1 with the $\gamma$-ray
emission. One way of confirming this association is to compare the
observed $\gamma$-ray emission structure with the CO gas, which is a
well-known tracer of the $\gamma$-ray emission zones, apart from
tracing the spiral arms better than HI in the inner disk (e.g.
\cite{damThad08}).

In order to study the Wd1 region in CO, we need to establish the
velocity of the Wd1 cluster. \citet{K&D07}, analyzed the neutral gas
environment around the cluster and concluded that Wd1 lies at a
radial velocity of $-50$~km~s$^{-1}$, thus locating it in the
external part of the Scutum-Crux (SCx) arm. Recent radial velocity
measurements of stellar members of Wd1 yield a mean velocity of
$-50~$km~s$^{-1}$ \citep{2009Ap&SS.324..321M, ritchie09}, thus
confirming the velocity derived from the HI analysis.

We have searched for molecular clouds towards Wd1, in the
Columbia-Calan CO survey data cube \citep{bronf89}. This survey
covers the entire southern MW and has a spatial resolution of
0.125$^\circ$, a velocity resolution of 1~km\,s$^{-1}$ and a
sensitivity of 0.1~K. Our analysis covers large angular scale
($2^\circ\times2^\circ$ centered at l=339.55$^{\circ}$,
b=$-0.4^{\circ}$), allowing us to study the large scale structure of
the interstellar medium around the cluster.

The aim of the present work is to search for the molecular clouds
associated with the H.E.S.S. $\gamma$-ray emission, and to
investigate whether Wd1 is the source of the observed $\gamma$-ray
emission. With this aim, we first analyzed the morphology and
kinematics of the molecular gas at the established radial velocity
of Wd1 ($-50$~km\,s$^{-1}$) in the SCx arm. The resulting poor
association of the CO structures with the H.E.S.S map, prompted us
to carry out the analysis of the CO morphology of the Norma arm at
$-90$~km\,s$^{-1}$, which is the next major arm in the line of sight
(LOS) to Wd1. In this latter arm, we discovered a ring, which could
be interpreted as an expanding shell, with very good structural
correspondence with the observed $\gamma$-ray emission. In this {\it
Letter}, we compare the CO maps at both velocities with the
$\gamma$-ray observations, and discuss in some detail the CO
structure of the Norma arm. In \S2, the CO molecular emission in the
direction of Wd1 is presented. The association between molecular gas
and VHE emission is presented in \S3, and finally in \S4, we discuss
the consequences of this correspondence.

\begin{figure}
\plotone{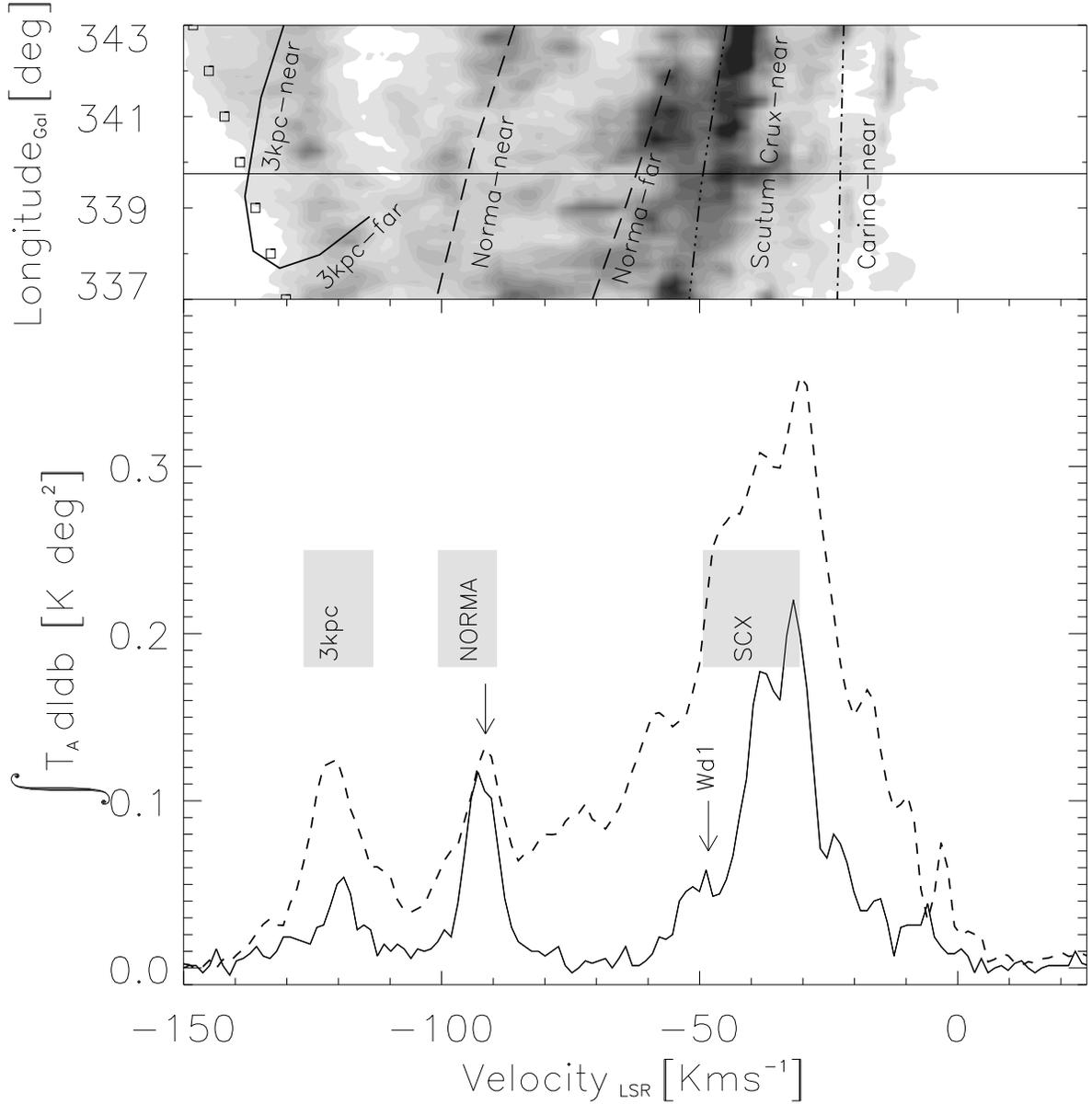} \caption{{\bf Top:} Reproduction
of the longitude-velocity diagram of \citet{bronf00} indicating the
position of spiral arms along at l=339.75$^\circ$ (solid line). The
flat rotation curve are marked with small squares. {\bf Bottom:} The
$^{12}$CO emission profiles in the direction of Wd1 in a narrow
($15^\prime\times15^\prime$; solid line), and wide
($2^\circ\times2^\circ$; dashed line scaled by a factor of 1/30)
beams. The range of radial velocities for 3 of the arms are
indicated by the shaded boxes. The arrows point to the radial
velocities obtained by fitting Gaussian profiles to the CO emission
peaks relevant to this study.}
\end{figure}

\section{$^{12}$CO emission in the general direction of Wd1}

The $^{12}$CO intensity line profiles along the LOS to Wd1 in narrow
and wide beams are shown in Fig.~1. This LOS intersects several arms
of the MW. We adopted the Galactic model of \citet{bronf00} to
identify the observed peaks with known spiral arms. The 3~kpc arm at
$-120.6$~km\,s$^{-1}$ with a FWHM of 13.7~km\,s$^{-1}$
(distance$\sim$7.5~kpc), is identified. The peak at
$-91.5$~km\,s$^{-1}$ with a FWHM of 8~km\,s$^{-1}$ corresponds to
the Norma near arm (distance=5.5$\pm$0.5~kpc), whereas the strongest
emission profile in the plot originates principally from the SCx arm
at $-35$~km\,s$^{-1}$ (distance$\sim3.5$~kpc). However, velocity
profiles of Norma far ($-50$~km\,s$^{-1}$, distance$\sim12.1$~kpc),
and Carina arm ($-10$~km\,s$^{-1}$, distance$\sim$0.5~kpc), also
contribute to the observed profile as can be inferred from Fig.~5 of
\citet{bronf00}. The far side of all arms in the fourth quadrant of
the MW, is very difficult to detect and/or distinguish, because they
are expected to be much fainter than the superposed contributions
from nearer arms. Hence, we assign the entire intensity of a
velocity peak to the nearer arm. We identified the peak located at a
velocity of $-48.3$~km\,s$^{-1}$ (FWHM of 14~km\,s$^{-1}$) with the
Wd1 cluster, following the HI study by \citet{K&D07}. Given that
they used a galactocentric distance to the Sun of 7.6~kpc, we
recalculated the distance to Wd1 using the IAU recommended value of
8.5~kpc, obtaining a distance of 4.3~kpc (instead of 3.9~kpc).

\begin{figure}
\epsscale{0.9}
\plotone{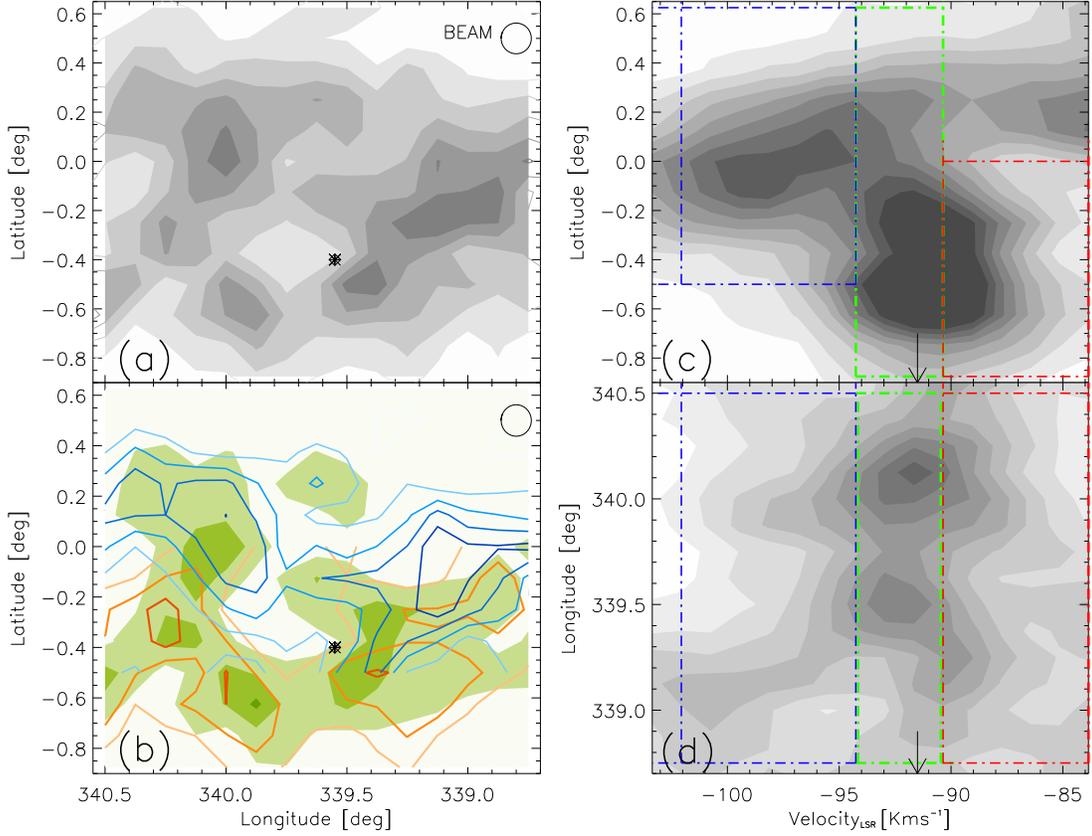} \caption{2D projections of the
$^{12}$CO data cube ascribed to the Norma arm ($-104<V_{\rm
rad}<-84$~km\,s$^{-1}$). {\bf (a)} Velocity integrated intensity
map. The lowest gray-scale level corresponds to $7\sigma$ above the
background, with successive levels increasing in steps of $7\sigma$.
{\bf (b)} Contour maps showing the receding (red) and approaching
(blue) gas with respect to that at rest ($-94<V_{\rm
rad}<-90$~km\,s$^{-1}$; green). The position of Wd1 is indicated by
an asterisk. {\bf (c)} Latitude-velocity map, and {\bf (d)}
Longitude-velocity map. An area of $1.5^\circ\times1.5^\circ$ is
integrated in the projected spatial axis in the last two panels,
where we also identify the slices of receding, approaching, and rest
gas with red, blue and green boxes, respectively. The cloud at
l=338.5 and b=0.1 (top-right box in panel c), has been excluded from
further analysis. The systemic velocity of the Norma arm is marked
with an arrow. }
\end{figure}

\begin{figure}
\plotone{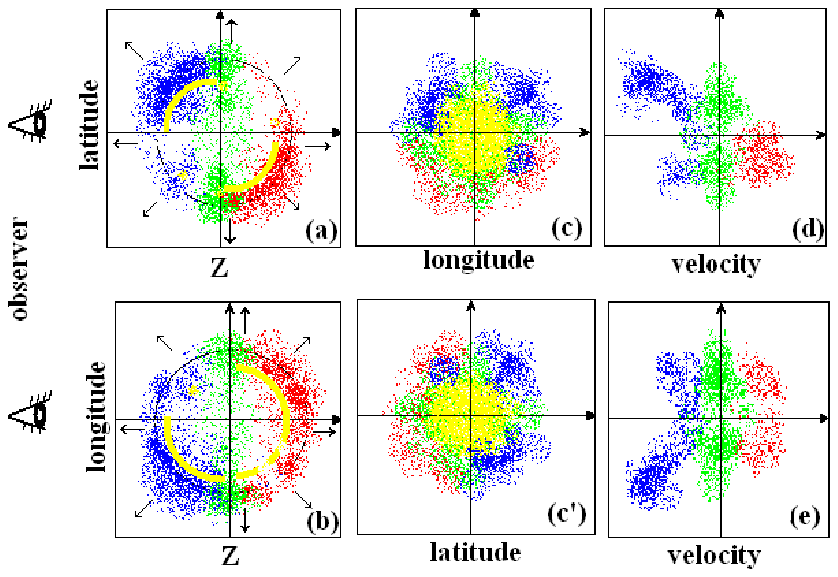} \caption{A proposed schematic
model based on the CO observation that reproduces the integrated
maps. Colors identifies the receding, approaching, and rest gas with
red, blue and green respectively. Z is the depth along the line of
sight. The $\gamma$-rays emission zone is marked with yellow and is
discussed in $\S3$).}
\end{figure}

The CO emission map centered at the radial velocity of the Norma arm
is shown in Fig.~2. In the velocity integrated map, an apparent CO
ring centered at l=339.75$^{\circ}$, b=$-0.4^{\circ}$, is easily
identifiable. This ring like structure constitutes a cavity
surrounded by a molecular gas complex of an outer diameter of
1.5$^\circ$, and a thickness with an average value of about
0.5$^{\circ}$. It is part of a fragmented shell seen in projection
(see below).

It can be seen in both position-velocity maps (Fig.~2c and 2d) that
there is both receding and approaching gas with respect to the
systemic velocity of the Norma arm. The observed CO emission largely
originates from the gas at the arm's systemic velocity. The
approaching gas is principally concentrated to the north of the ring
center, whereas the receding gas is located to the south. The
approaching gas is not concentrated in a single structure, instead
the observed emission comes mainly from two regions, one located at
l=340.0$^{\circ}$, b=0.0$^{\circ}$ and the other at
l=339.2$^{\circ}$, b=0.0$^{\circ}$. They have a maximum radial
velocity of 8~km\,s$^{-1}$ with respect to the systemic velocity
($-91.5$~km\,s$^{-1}$) of the Norma arm. These components are
compact in projection and they merge in velocity smoothly with the
Norma's systemic gas, as can be seen in fig.~2(d). This leads us to
believe, that the approaching component belongs to the Norma arm,
and not to the 3~kpc far arm. The receding component is distributed
uniformly along the latitude and longitude range of the cloud with a
maximum velocities of 6~km\,s$^{-1}$ with respect to the arm's
systemic velocity. Fig.~2b shows these velocity components
superposed on the spatial map. All these observed characteristics
suggest that the molecular ring is expanding at 6 to 8~km\,s$^{-1}$.

In Fig.~3, we present 2D sections of a 3D expanding shell schematic
model that best explains the observed structures in space and
velocity. In Fig.~3 (a),(b),(c) and (c') we show the projections in
latitude-depth, longitude-depth and latitude-longitude planes,
respectively. In each panel the approaching and receding gas is
shown in blue and red colors respectively, whereas the gas at rest
is shown in green. The observed spatial distribution of gas, as well
as the observed kinematics, are well reproduced by our simple
schematic model, in which the gas principally resides on the surface
of an expanding shell, with very little gas in the center of the
shell. The gas in the shell is not uniformly distributed, instead
the majority of the gas mass is located in two diagonally opposite
directions. From the model, we infer that the gas at rest (in green)
is also participating in the expansion, but it is seen at rest due
to projection effects. In Fig.~3 (d) and (e), we show the
projections in the latitude-velocity and longitude-velocity planes
respectively. These figures reproduce well the observed structures
in Fig.~2 (c),(c') and (d), respectively.

We estimate the molecular cloud mass content in the shell to be
10$^6$M$_{\odot}$. For calculating this, we used the ratio
N(H$_2$)/W$_{^{12}CO}$=
1.56$\times$10$^{20}$\,cm$^{-2}$\,K$^{-1}$\,km\,s$^{-1}$
\citep{hunt97}, and with the standard approximation
M$_{cloud}$=(2)\,(1.36)\,A\,m$_p$\,N(H$_2$), where A is the
projected area, m$_p$ the proton mass, N(H$_2$) the molecular column
density, and W$_{CO}$ the $^{12}$CO integrated intensity, the factor
1.36 accounts for Helium assuming a 10$\%$ abundance by number (See
Murphy $\&$ May 1991). The kinetic energy involved in the observed
gas mass and expansion velocities is about $\sim10^{51}$~erg using
E$_{kin}$=(1/2)\,M$_{cloud}$\,v$_{exp}^2$. The observed size of the
shell implies that the cavity was produced around 6~Myr ago, if the
expansion velocity was uniform over time, or shorter if the shell
had a larger expansion velocity in the past.

A map showing the distribution and the kinematics of the atomic HI
gas associated with the molecular cavity is presented in Fig.~4 of
\citet{luna2009}. We note that the distribution and kinematics of
the atomic HI gas is very similar to that of the CO emission. The
cavity observed in the molecular gas is seen in the atomic gas as an
HI hole of low column density. The lowest HI integrated emission
outlines the northern inner boundary of the molecular cavity.

\section{The molecular gas associated to the $\gamma$-ray emission}

\begin{figure}
\epsscale{0.6} \plotone{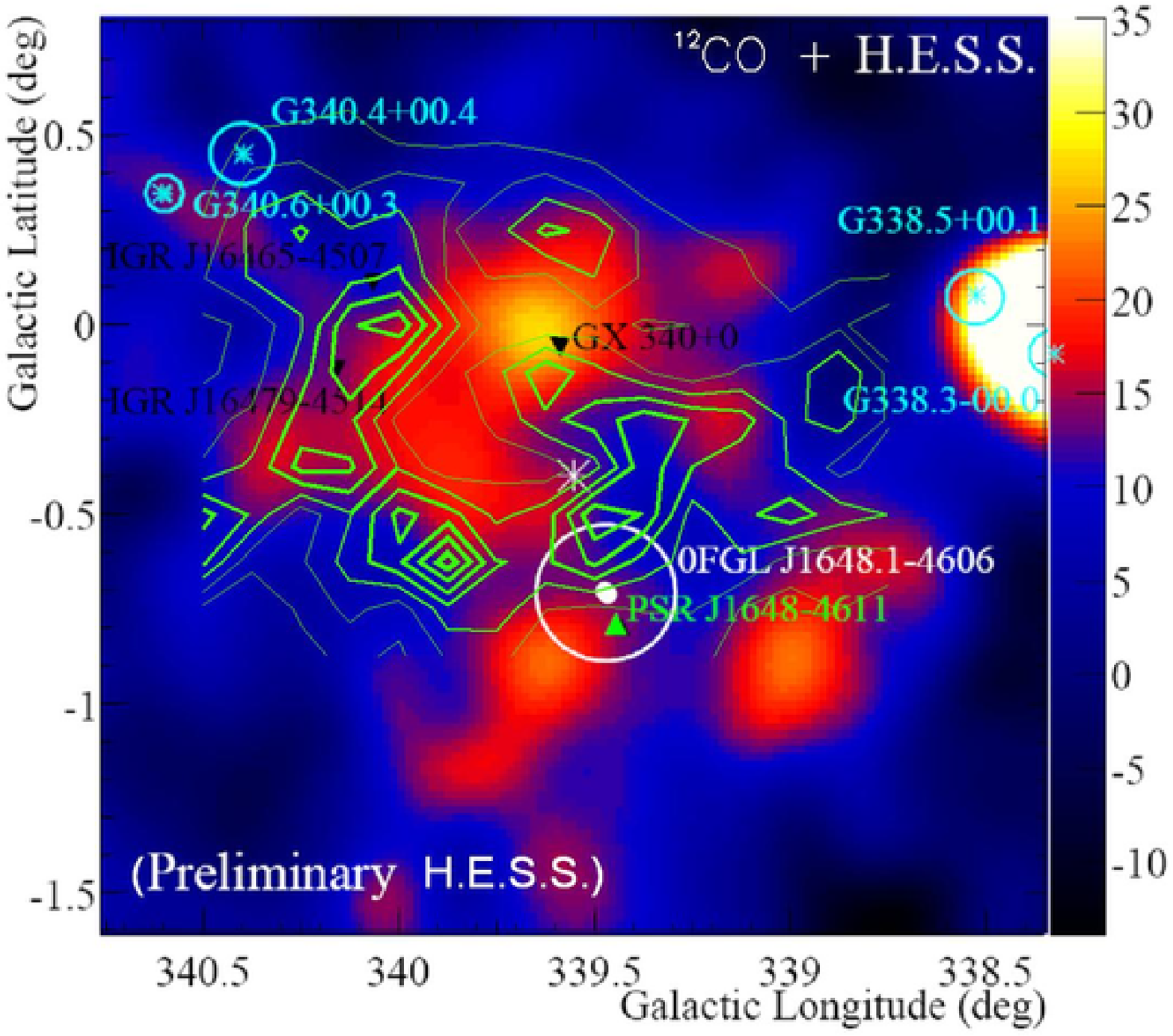} \plotone{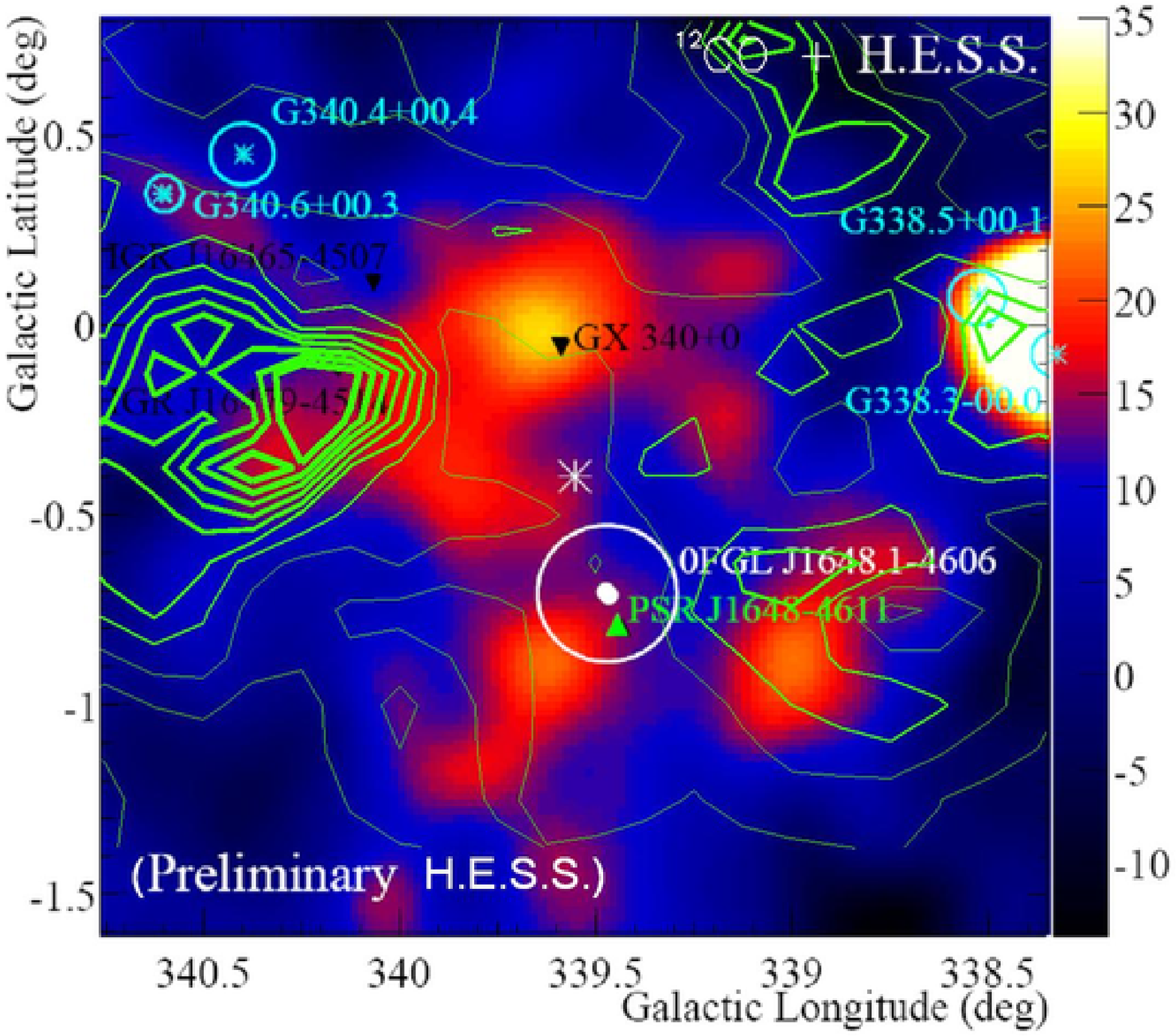} \caption{ The
extended VHE $\gamma$-ray emission detected by the H.E.S.S.
collaboration (Ohm09, their ``preliminary" Fig. 3), where we have
overplotted green contours of integrated CO gas at rest with respect
to Norma arm ($V_{\rm rad}$=$-$91~km\,s$^{-1}$,
$\Delta$V=5~km\,s$^{-1}$ every $3\sigma$ from $3\sigma$)({\bf top}),
and CO gas at rest with respect to Wd1 ($V_{\rm
rad}$=$-$48~km\,s$^{-1}$, $\Delta$V=20~km\,s$^{-1}$ every $10\sigma$
from $10\sigma$)({\bf bottom}). Thickness of the line contours are
decreased from bright to faint levels to remark structures.}
\end{figure}

In this section, we compare the morphology of the molecular cavity
with the $\gamma$-ray emission map from Ohm09. The top panel of
Fig.~4 shows the CO emission associated with the molecular shell in
contours superposed on the $\gamma$-ray emission map, which has a
beam of $30^\prime$. The latter map is dominated by several bright
point sources that have been identified by Ohm09 as associated with
already known sources. The X-ray source GX340+0 at l=339.56,
b=$-$0.085, the pulsar PSR J1648-4611, and the magnetar CXO
J164710.2-455216 (white asterisk in Fig.~4), are of relevance to the
present paper. In addition to these point sources, there is extended
emission in the shape of a horseshoe, that coincides with the
projected inner wall of the molecular shell. Especially noticeable
is the similarity of the structures of the $\gamma$-ray emission and
the dense molecular gas northward of the magnetar, where the
projection of the molecular shell is seen as a protrusion inside the
cavity. The $\gamma$-ray emission seems to fill the cavity in the
figure. Such a configuration can be produced in our 3D expanding
shell model, if the $\gamma$-ray emission comes from the inner
surface of the molecular shell as can be seen in yellow shades in
Fig.~3. Hence, the proposed shell model with the $\gamma$-ray
emission restricted to the inner wall of the shell, can explain not
only the observed molecular position-velocity map, but also the
$\gamma$-ray emission structure observed by H.E.S.S.

While the molecular shell of the Norma arm could be traced in many
successive channel maps. Yet, there is no coordinated structure in
anyone the velocity channel maps around the systemic velocity of Wd1
($\approx-$50~km\,s$^{-1}$). Hence, in order to generate a CO map of
the Wd1 region, we integrated all the channels in a 20~km\,s$^{-1}$
velocity range centered at $-$48~km\,s$^{-1}$. In the bottom panel
of Fig.~4, we show this CO map superposed on the $\gamma$-ray
emission map. There is no coordinated CO structure associated with
Wd1, that could be compared with the $\gamma$-ray emission map. The
only prominent structures seen in CO are at l=340.25$^{\circ}$,
b=$-0.30^{\circ}$ and l=339$^{\circ}$, b=0.5$^{\circ}$. Hence, Wd1
is not responsible for the observed extended $\gamma$-ray emission.

\section{Summary and Discussions}

We analyzed the morphology and kinematics of the molecular
environment in the general direction of Wd1, with the aim of looking
for the molecular gas that may be associated with the recently
reported H.E.S.S. $\gamma$-ray emission.  We discovered an expanding
molecular ring at the radial velocity of $-91.5$~km\,s$^{-1}$, with
the inner contours of the ring coincident with the observed
$\gamma$-ray emission structure. The observed morphologies of the
$\gamma$-ray emission and the CO emission along with their
kinematics, can be understood in terms of an expanding fragmented
shell of molecular gas. Thus, the 2 components are physically
associated and both lie in the Norma arm. Consequently, the
molecular structure surrounding the Wd1 ($-50$~km\,s$^{-1}$) cluster
does not correspond to the observed $\gamma$-ray emission map.

The $\gamma$-ray emission is produced by accelerated energetic
particles when they interact with dense gas. Recently,
\citet{fujita09} proposed a model of generating VHE $\gamma$-rays by
a SNR located in a cavity surrounded by high density molecular gas
in order to explain the observed photon spectra for the hidden SNR
in the open cluster Wd2, and the old-age mixed-morphology SNR W28.
In this model, the particles are accelerated at the end of the Sedov
phase, when the SN shock reaches and collides with the surrounding
spherical high-density molecular gas (shell). This interaction
produces $\gamma$-ray emission in the inner wall of the molecular
shell, exactly the configuration that produces our observed CO
morphology. Hence, the presence of a stellar remnant inside the
molecular cavity can give rise to the observed $\gamma$-ray
emission. There are three stellar remnant candidates in the field of
view. They are a binary system GX340+0 containing a neutron star
\citep{schultz93}, a magnetar (CXO J164710.2-455216), and the pulsar
PSR J1648-4611, with the third one the most likely precursor.

The first two sources are seen geometrically inside the molecular
cavity, but lie at distances different from that of the cavity
according to their present distance estimations. The magnetar is
associated with Wd1 following the analysis by \citet{muno06}, based
on statistical grounds. However, it could as well be located further
out, in the molecular cavity. For the neutron star in the binary
system GX340+0, a maximum distance of 11$\pm$3~kpc has been
estimated based on its luminosity, assuming that it is radiating at
the Eddington limit \citep{penninx93}. Yet, it is conceivable that
the neutron star is not emitting with the limiting luminosity, in
which case it could be located at the distance of our cavity.

A third stellar remnant (PSR J1648-4611) is located half a degree
South-East of the ring center and is seen just at the outer border
of the molecular ring (see Fig.~4). A distance by dispersion measure
yields a value of 5--5.7~kpc, depending on the electron density
distribution model adopted \citep{Torres03}. That is at the same
distance of the cavity. This makes this pulsar a strong candidate
for producing the $\gamma$-ray emission at the inner boundaries of
the molecular shell. The location of the pulsar off-center of the
cavity, could be due to a large proper motion of the pulsar.
following the SN explosion. It requires only a velocity as little as
8~km\,s$^{-1}$ for the pulsar to move from the center of the shell
to its present position, if the explosion occurred 6~Myr ago, and
the pulsar is moving in the plane of the sky.

Recently, the LAT instrument on board of the Fermi satellite
detected unpulsed emission from a region coincident with this pulsar
\citep{abdo09}. The model proposed by \citet{fujita09} expects
emission in the Fermi band coincident with the $\gamma$-ray
emission, but such emission is not yet detected by Fermi
observations. However, the observed configuration of a cavity
surrounded by a molecular shell, with associated VHE $\gamma$-rays
emission, and a bright Fermi source in the vicinity, are common to
both, Wd2 and the expanding shell discussed in our work.

\acknowledgments L.B. acknowledges partial support from Centro de
Astrof\'isica FONDAP 15010003 and from Center of Excellence in
Astrophysics and Associated Technologies PFB 06. The authors are
grateful to an anonymous referee whose comments and suggestions have
largely improved the clarity of this paper.

\end{document}